\documentstyle[preprint,aps]{revtex}
\def\ra{\rightarrow}
\begin{document}
\preprint{\vbox{\baselineskip=14pt
  \rightline{UH-511-946-99} \break
  \rightline{November 19, 1999} \break
        }}
\tightenlines
\draft
\title{Relativistic Quantum Computing}
\vspace{.25 in}
\author{S. Pakvasa, W. Simmons  \& X. Tata}
\vspace{.75 in}
\address{Department of Physics \& Astronomy \\
 University of Hawaii at Manoa \\
              Honolulu, HI 96822}


\maketitle

\vspace{.75 in}

\abstract
{We present some informal remarks on aspects of relativistic quantum computing.}
\vspace{.50in}

\vspace{.10 in}

\indent	
The continuing miniaturization of electronic circuitry has forced designers of 
computer chips to deal with all of the complications that are entailed in using 
Quantum Mechanics.  Meanwhile, physicists have identified dramatic new 
possibilities for increases in computer performance by making use of such 
uniquely quantum effects as superposition, the entanglement of states, and the 
collapse of wave functions during measurement.  Quantum computation and 
quantum information theory have now become exciting new fields of research.  As 
circuits become yet smaller and the demand for higher clock speeds continues to 
escalate, relativistic effects, as well as quantum effects, will become important in 
computing.

\vspace{.10 in}

\indent	
The energy-time uncertainty relation in Quantum Mechanics places a 
number of interesting bounds on the performance of computers.  We shall show 
that as faster computer circuits are developed, a series of thresholds is encountered, 
each requiring major changes of engineering approach.  Eventually, in order to 
design the very fastest computer circuits, engineers will be forced to deal with the 
full machinery of Relativistic Quantum Field Theory, which is today used mainly 
by high energy physicists who study relativistic particle collisions.

\vspace{.10 in}

\indent	
Let us pause to think back to the level of technology available to engineers 
in the early part of the twentieth century at the time that Einstein proposed his 
theory of the specific heat of solids.  Just after his theory was proposed, the internal 
states of solids were known to be astronomical in number, multi-particle, 
quantized, and generally evanescent in character.  Those internal states could be 
manipulated only in the crudest way with the equipment available at that time.  
Physicists and engineers of that period evidently did not anticipate that by the last 
quarter of the century, the control of internal states of materials would be at the 
core of all modern technology.  In fact, given some of the well known and 
laughable negative predictions made by physicists at various times past, it seems 
possible that many physicists would have dismissed out of hand the possibility of 
electronic micro-circuitry.

\vspace{.10 in}

\indent	
	Today we know that relativistic collisions produce complex, non-linear, 
multi-particle, quantized states that are highly evanescent and individually 
inaccessible to our equipment. We suggest that is not unreasonable to suppose that 
one day these relativistic states will be technologically significant in new ways.

\vspace{.10 in}
{\noindent{\underline{Uncertainty Principle Limitations}}}

\vspace{.10 in}

\indent		
It is common in classical computer engineering to associate a bit with an 
energy state.  For example, a 1 might be represented by a high voltage and a 0 by a 
low voltage.  As long as it is necessary to distinguish amongst states
with distinct values of 
energy in order to perform logical operations, then the energy-time uncertainty 
relation dictates that there be a minimum energy associated with any given clock 
rate.  For a clock frequency, $f$, the minimum energy, $\Delta E$, required by the 
uncertainty relation is, (in atomic units),
\begin{equation}
f= \Delta E
\end{equation}
 		
	Specifically, one electron-volt corresponds to a frequency of 
$1.5 \times 10^{15}$ Hertz, or 1.5 peta-Hertz\footnote{Peta-Hertz equals 
$10^{15}$ Hertz, Exa-Hertz equals $10^{18}$ Hertz, Zetta-Hertz 
equals $10^{21}$ Hertz, and Yotta-Hertz 
equals $10^{24}$ Hertz.

Femto-second equals $10^{-15}s$,
Atto-second equals $10^{-18}s$, 
Zepto-second equals $10^{-21}s$, 
and Yocto-second equals $10^{-24}s$.}, which is a million-fold faster than 
today's machines.  

\vspace{.10 in}

\indent	
	This relationship is quite general and applies to any form of energy when 
used to distinguish bits; light, electric charge separation, electron spin in a 
magnetic field, etc.  Moreover, it applies to any mechanism that uses energy to 
distinguish bits, whether it be a CPU, a memory device, or communication line.

\vspace{.10 in}

\indent	
	For the same reasons, the power flowing through a logic circuit like the ones 
described above, must increase with the square of the clock frequency.  We note 
that nothing dictates that the power flowing through the circuit must be wasted, but 
it must flow for the circuit to do its job within the restrictions of the uncertainty 
relation, and this quadratic increase of power with frequency will become very 
burdensome. 
\begin{equation}
\mbox{Power Per Bit} \ra  f^2 
\end{equation}

As we shall discuss, below, the two simple and fundamental, equations (1) 
and (2), when taken together with the properties of matter and radiation, determine 
the shape of the future of both classical and quantum computing.

\vspace{.10 in}

\indent	
	A quantum circuit that processes an input string, would read in qubits, which 
are two component superposition states, one by one. Suppose that the engineering 
choice is made to limit each qubit to only one of two energy eigenstates, which 
differ in energy by  E.  That is, each qubit carries only one bit of information.  In 
this case, Eq. (1) gives us the maximum frequency of this particular quantum 
circuit.

\vspace{.10 in}

\indent	
	It has been known for more than twenty years that a single qubit cannot 
carry more information than the single bit just discussed.  For example, if we set 
the qubit to be a superposition of the two eigenstates, then there are up to three real 
parameters that describe the mixing.  But a single measurement of the qubit will 
result in one or the other of the eigenvalues and can yield essentially no 
information about the mixing angles.

\vspace{.10 in}
{\noindent{\underline{The Future of Computers}}}

\vspace{.10 in}

\indent	
	With the limitations imposed upon computers by the uncertainty relation in 
mind, let us explore the future of this technology.

\vspace{.10 in}

\indent	
	Let us begin with a simple application of Eq. (1).  If all of the 
properties of semiconductors, which make them so useful, are to be available, then 
the energy of the electron state should be less than the energy gap, which might 
typically be one electron-volt.  This implies the existence of a sort of barrier at the 
frequency corresponding to one electron-volt, i.e. 1.5 peta-Hertz.  Above that 
frequency, unique semiconductor properties start to become unavailable.  This 
barrier, like the sound barrier, is in some sense illusory, entailing no problem of 
principle.  Surpassing the barrier simply requires changing the technology.

\vspace{.10 in}

\indent	
	As another example, suppose that bits are distinguished using the energy of 
orientation of a spin relative to a magnetic field.  Then for a given frequency, the 
magnetic field strength must be large enough so that the magnetic spin energy, 
(which is proportional to the magnetic field strength) is at least as large as the 
frequency.  It follows that the energy stored in the magnetic field, (which goes like 
the square of the field strength), must increase with the square of the frequency, as 
does the power passing through the spin states.  Thus, in this example, both the 
(stored) energy of the computer mechanism and the power passing through its 
circuits increase like the square of the frequency.

\vspace{.10 in}

\indent	
	As a part of the trend toward higher clock frequencies, circuit engineers are 
reducing the sizes of the elemental circuit elements.  This means that they must 
deal with individual electrons, whose energies are increasing with the frequency.  
The smaller circuits also encounter all of the well known problems associated with 
controlling quantum particles.

\vspace{.10 in}

\indent	
	As we have said, as the desired clock frequency increases, the minimum 
energy required to represent information increases in proportion.  However, 
quantized systems will, in general, have energy levels spaced closer than this 
minimum, thus and these levels cannot be used to distinguish information.  Thus, 
as clock speed increases, there is a decrease in the number of energy states 
available for representing information.  For example, the hydrogen atom has an 
infinite number of levels but only some of them could be used to store information 
if one wants to get out answers in a finite time.  One possible way to release more 
levels for use is to engineer the over-all energy of the circuit elements yet higher; 
that is, choose quantum systems of greater total energy.

\vspace{.10 in}

\indent		
As a final example of the importance of the uncertainty relation in 
computing, we note that it has recently been suggested that spin-spin interactions 
in semiconductors may be useful in quantum computing.  If spin domains are used 
to store bits, then the clock period will be limited by the light travel time across the 
domains.  If individual spin-spin interactions are used, in the interest of speed, then 
the energy of interaction is of the order of $10^{-4}s$ eV 
and the limiting frequency is only about 150 GHz.
\vspace{.10 in}

{\noindent{\underline{Relativistic Effects}}}

\vspace{.10 in}

\indent	
	As engineers respond to the limitations discussed above, they must push the 
maximum operating energy of circuit elements higher in order to facilitate higher 
clock frequencies demanded by users.  If the rapid progress in clock speed 
proceeds as in recent years, then relativistic effects will appear shortly.  Once the 
limitations entailed in the uncertainty relation become important, then, as 
mentioned above, the power throughput will increase quadratically with frequency.  
Therefore, computers of the future may become high power, relativistic devices 
similar to particle accelerators.

\vspace{.10 in}

\indent	
While, engineering practice has included classical relativistic electronic 
effects ever since the introduction of radar, the combination of relativistic and 
quantum effects in logic circuits will probably come as a rude surprise to engineers 
who will have to begin using relativistic quantum mechanics or quantum field 
theory

\vspace{.10 in}

\indent	
One relativistic effect, which will ultimately turn out to be very important in 
computing, is the impossibility of localizing an electron to a volume smaller than 
that characterized by electron's Compton wavelength.  An important implication of 
this, which is discussed below, is that the time required for light to cross this small 
distance, which is about one zepto-second, 
represents a minimum time for a logical process involving electrons, and a 
maximum operating frequency measured in zetta-Hertz.  Furthermore, at zetta-
Hertz frequencies, there is another significant relativistic effect. At a frequency of 
a few zetta-Hertz the energy-time relationship implies that the voltage in the single 
electron logic circuits will exceed the threshold for the production of electron-
positron pairs.  The positrons will quickly annihilate with environmental electrons, 
producing 0.5 MeV gamma rays.  

\vspace{.10 in}

\indent	
	In semiconductor circuits, the annihilation of electron-hole pairs produces 
electromagnetic radiation and that has not inhibited the development of practical 
circuits - in fact, the radiation is often quite useful.  Similarly, the production of 
gamma rays may not inhibit the development of high energy circuits.  

\vspace{.10 in}

\indent	
	The limitation upon the localization of electrons is, however, a barrier of 
principle.  Present day concepts of computer circuitry, based, as they are on 
electrons, and extrapolated to high frequency and single particle circuits, simply 
will not work above the zetta-Hertz frequency range.  
\vspace{.10 in}

{\noindent{{\underline{Power Requirements}}}

\vspace{.10 in}

\indent	
	Based upon the discussion so far, we can estimate, very roughly, the power 
consumption of a typical relativistic quantum computer.  

\vspace{.10 in}

\indent	
	Let us suppose we want to have a machine that operates on a clock cycle of 
one zepto-second and which represents each bit by a single electron.  From the 
uncertainty relation, the electron kinetic energy must be at least one MeV.  If the 
through-put is one kilo-bit wide, then the machine has a throughput of one yotta-bit 
per second. This might be 
a nice piece of equipment to look forward to owning, because it is about fourteen 
orders of magnitude faster than the 32 bit, 500 MHz machine that may now 
be sitting on the reader's desk. (Note, for example, that a computation that takes 
just one second at a rate of one yotta-bit/second would require thirty million years 
to complete on your desk machine.)

\vspace{.10 in}

\indent	
	For simplicity in estimating the power requirements, let us set aside the 
possibility of reversibility and of recycling energy from cycle to cycle. The 
elementary circuit process involves energizing one thousand electrons and making 
a logical operation on them in one clock cycle, then repeating the process during 
the next cycle.  At the uncertainty limit, the power is just the number of electrons 
operating in parallel, times the energy of each electron, times the frequency. The 
power required in this example is about one hundred billion watts!  Considering 
that almost all of that comes out of the machine as gamma rays, it might not be a 
desirable replacement for the machine on your desk, after all.

\vspace{.10 in}

\indent	
The power estimated, above, is only the power for a single tier of the most 
elementary circuit elements.  To estimate the power of the entire computer, the 
power, above, must be multiplied by the number of elementary circuit elements in 
the machine, which is generally a large number.
\vspace{.10 in}

{\noindent{{\underline{Ultimate Speed Computer}}}

\vspace{.10 in}

\indent	
	The question naturally arises as to whether there is any ultimate physics 
limit to the speed of logic circuitry, or whether we can build as fast a computer as 
we like if we are clever enough and are willing to commit unlimited power.  There 
is no fundamental unit of time in physics.  That implies that there is no known 
upper limit to how fast an event can take place.  However, at a fundamental level, 
there is a fastest process.  

\vspace{.10 in}

\indent	
	The fastest fundamental process now known to physics, is the formation or 
disintegration of the $Z^0$ boson.  The lifetime of the $Z^0$ is about one 
hundredth of a yocto-second, which 
corresponds to a frequency of one hundred yotta-Hertz.  Therefore, assuming that 
no circuit can operate faster than the fastest fundamental process, there is an 
ultimate limit to computation speed.  However, as emphasized earlier, due to 
relativistic effects, electrons cannot be measured in times less than about one 
zepto-second.  Therefore the ultimate clock speed for an electronic circuit would 
be in the range of zetta-Hertz, which was used in the example in the previous 
section.

\vspace{.10 in}

\indent	
	These same limitations apply to digital communications.  The fastest 
processes that can be sustained using electrons are in the zetta-Hertz range.
\vspace{.10 in}

{\noindent{{\underline{How Soon Will We Reach These Limits?}}}

\vspace{.10 in}

\indent	
	Computer clock speed has been increasing dramatically in recent decades 
and may well fit the exponential progress model.  Early in this century data 
communication was accomplished by using Morse code and by using manual 
keyboard entry into hand cranked mechanical calculators.  Today, the same 
functions are performed by giga-Hertz or tera-Hertz data communication lines.  
This represents an increase of ten or twelve orders of magnitude in speed over 
eight or nine decades.  For discussion purposes, let us take the average figure to be 
an order of magnitude increase in clock speed, or equivalently, operating 
frequency, every ten years.  Is it plausible to suppose that progress in computer 
speed will continue at this pace?

\vspace{.10 in}

\indent	
	At each stage of development, the newest computers are applied to the 
problem of designing better and faster equipment and for improving and 
controlling the manufacturing processes.  For example, enormous progress has 
been made in recent years in simulating the properties of materials on the 
computer; and this accelerates the development of materials to be used in the 
manufacture of better equipment.  At each stage of computer development, the 
existing computers provide both tools for research and design and provide 
components for yet more sophisticated machines.  Additionally, all of the new 
applications have opened up new markets causing an influx of capital into 
development of new technology.  It is thus, not implausible to speculate that the 
rate of increase of clock speed is proportional to clock speed.

\vspace{.10 in}

\indent	
Let us suppose, then, that computer clock speeds are in an exponentially 
increasing stage of development, and assume that the same pace, of an order of 
magnitude per decade, will continue indefinitely.  From today's giga-Hertz 
machines to peta-Hertz machines will take six decades.  That is about all of the 
time available to use today's semiconductor and quantum dot technology.
With the same extrapolation in 
another six decades, around the year 2120, the zetta-Hertz barrier will be 
encountered.  In some sense, that will mark the end of the era of electronics.
\vspace{.10 in}

{\noindent{\underline{Conclusions}}}

\vspace{.10 in}

\indent	
	Quantum Mechanics, together with elementary properties of matter and 
radiation, determine the future of computers and computation.  The energy-time 
uncertainty relation, together with strong engineering trends, implies 
the ultimate computer will have to incorporate the principles of 
relativistic quantum physics.
\vspace{.10 in}

\indent	 
This work is supported in part by a grant from U.S. Department of
Energy.

\vspace{.10 in}
	     
\noindent
{\underline{Note Added:}} After completing this work, we became aware of a paper by
Lloyd \cite{lloyd}, which also addresses physics limitations of the
ultimate computer.

\end{document}